\documentclass[prd,aps,preprint,nofootinbib]{revtex4}
\usepackage{epsfig}
\usepackage{amsmath}

\begin{document}

%\preprint{LBNL-xxxx} \preprint{RBRC-xxx}

\title{Hyperon Polarization in Unpolarized Scattering Processes}

\author{Jian Zhou}
\email{jzhou@lbl.gov} \affiliation{ Nuclear Science Division,
Lawrence Berkeley National Laboratory, Berkeley, CA
94720}\affiliation{School of Physics, Shandong University, Jinan,
Shandong 250100, China}
\author{Feng Yuan}
\email{fyuan@lbl.gov} \affiliation{Nuclear Science Division,
Lawrence Berkeley National Laboratory, Berkeley, CA
94720}\affiliation{RIKEN BNL Research Center, Building 510A,
Brookhaven National Laboratory, Upton, NY 11973}
\author{Zuo-Tang Liang}
\email{liang@sdu.edu.cn} \affiliation{School of Physics, Shandong
University, Jinan, Shandong 250100, China}
%\date{\today}

%\vspace{-1.5in}
%\vspace{1.4in}

\begin{abstract}
Transverse polarization in the Hyperon ($\Lambda$) production in the unpolarized
deep inelastic scattering and $pp$ collisions
is studied in the twist-three approach, considering the contribution
from the quark-gluon-antiquark correlation distribution in nucleon. We further compare
our results for deep inelastic scattering to a transverse momentum
dependent factorization approach, and find consistency between the two approaches
in the intermediate
transverse momentum region. We also find that in $pp$ collisions, there are only derivative
terms contributions, and the non-derivative terms vanish.
\end{abstract}

\maketitle

\section{Introduction}
Single transverse spin asymmetry (SSA) phenomena have a long history, starting with
the observation of large transverse polarization of Hyperon ($\Lambda$)
in unpolarized nucleon-nucleon scattering~\cite{bunce,review}. It has imposed theoretical
challenges to understand these phenomena in Quantum Chromodynamics (QCD)~\cite{review,qiu}.
In recent years, this physics has attracted strong interests from both experiment and theory sides.
For example, the experimental observation of SSAs in semi-inclusive
hadron production in deep inelastic scattering (SIDIS), in inclusive hadron
production in $pp$ scattering at collider energy at RHIC, and the
relevant azimuthal asymmetric distribution of hadron production in $e^+e^-$ annihilation
have motivated theoretical developments in the last few years~
\cite{Brodsky:2002cx,Collins:2002kn,Ji:2002aa,Boer:2003cm,JiMaYu04,{ColMet04}}.
Among these developments, two mechanisms in the QCD framework
have been most explored to study the large SSAs observed in the experiments.
One is the so-called twist-three quark-gluon correlation approach~
\cite{qiu,Efremov,Kanazawa:2000hz,new,Eguchi:2006qz,Eguchi:2006mc,{yuji-compact}},
and the other is the transverse momentum dependent (TMD) approach where the intrinsic
transverse momentum of partons inside nucleon plays important role
~\cite{Brodsky:2002cx,Collins:2002kn,Ji:2002aa,{Boer:2003cm},Sivers:1990fh,{Anselmino:1994tv}}.
Recent studies have shown that these two approaches are consistent to each other
in the intermediate transverse momentum region where both apply~\cite{Ji:2006ub}.

In particular, theoretical developments have been made to understand the
$\Lambda$ polarization in unpolarized hadronic reactions~\cite{goldstein,{Liang:1997rt},yuji-l,anselmino-l}.
Similar to the SSA in inclusive hadron production in $p^{\uparrow}p\to \pi X$, the
Hyperon polarization in unpolarized $pp$ scattering $pp\to \Lambda_{\uparrow} X$
receives the contributions from (naive)-time-reversal-odd effects in the distribution
and fragmentation parts~\cite{yuji-l,anselmino-l}. In order to understand
the experimental observations of $pp\to \Lambda_{\uparrow}X$,
one has to take into account both contributions.

On the other hand, in the deep inelastic scattering process,
one can separate these two contributions because they have different
azimuthal angle dependence~\cite{boer98}. To describe these effects, one can
calculate the $\Lambda$ polarization in the twist-three approach~\cite{yuji-l} or use
the TMD mechanism~\cite{boer98}. For example, the contribution from the T-odd
effects in the distribution part is associated with the so-called Boer-Mulders TMD quark
distribution $h_1^\perp$~\cite{boer98} multiplied by the TMD transversity fragmentation function $H_{1T}$
when the transverse momentum of the produced $\Lambda$
is much smaller than the hard scale $P_{\Lambda\perp}\ll Q$
where $Q$ is the virtuality of the virtual photon in DIS process. On the
other hand, we can also calculate this contribution from the
twist-three mechanism when the transverse momentum is much larger
than the nonperturbative scale $\Lambda_{\rm QCD}$: $P_{\Lambda\perp}\gg \Lambda_{\rm QCD}$.
From our calculations, we find that these two approaches indeed provide a unique description for
$\Lambda$ polarization at the intermediate transverse momentum region in the
semi-inclusive DIS. The large transverse momentum Boer-Mulders function
calculated in this context can also be used in other processes, like Drell-Yan
lepton pair production in $pp$ scattering~\cite{boer-dy}.

The rest of the paper is organized as follows.
In Sec.II, we study the $\Lambda$ polarization in
semi-inclusive deep inelastic scattering $ e+p
\rightarrow e + \Lambda_\uparrow +X $  by calculating the
contribution from the twist-three quark-gluon-antiquark correlation
function from nucleon. We will then take the
limit of small transverse momentum $P_{\Lambda\perp}\ll Q$,
and compare to the prediction from the TMD factorization
approach. In the TMD picture, the polarization of $\Lambda$
comes from the Beor-Mulders function $h_1^\perp$.
In Sec.III, we extend our calculations to the Hyperon polarization
in $pp$ collisions, and compare to the previous results.
We conclude in Sec. IV.

\section{$\Lambda$ transverse polarization in semi-inclusive DIS}

In the SIDIS process $ e p\rightarrow e' \Lambda_\uparrow X $ ,
 the differential cross section can be formulated as
\begin{eqnarray}
\frac{d\sigma (S_{\Lambda\perp})}{dx_B dy dz_h d^2\vec{P}_\Lambda }
=\frac{2 \pi \alpha_{em}^2}{Q^2} L^{\mu \nu }(l,q) W_{\mu \nu}(P,q,P_\Lambda,S_{\Lambda \perp})\ ,
\end{eqnarray}
where $\alpha_{em}$ is the electromagnetic coupling,
$l$ and $P$ are incoming momenta for the lepton and nucleon,
$q$ the momentum for the exchanged virtual photon with $Q^2=-q.q$,
$P_\Lambda$ and $S_{\Lambda\perp}$ are the momentum
and transverse polarization vector for the final state $\Lambda$, respectively, and
we have  $S_{\Lambda \perp} \cdot  P_\Lambda =0$. The kinematic variables are defined
as $x_B=\frac{Q^2}{2 P \cdot q}$, $z_h=\frac{P \cdot P_\Lambda}{P \cdot q}$,
$y=\frac{P \cdot q}{P \cdot l}$.
In the above equation,  $L^{\mu \nu}$ and $W^{\mu \nu }$ are the leptonic
and hadronic tensors, respectively. They are given by,
\begin{eqnarray}
L^{\mu \nu}(l,q)&=&2(l^\mu l^{'\nu}+l^\nu l^{'\mu}- g^{\mu \nu} \frac{Q^2}{2})\ ,\\
W^{\mu \nu}(P,q,P_\Lambda,S_{\Lambda \perp})&=&\frac{1}{4z_h} \sum_X
\int \frac{d^4\zeta}{(2 \pi)^4} e^{iq \cdot \zeta} \langle P |J_\mu(\zeta)|X P_\Lambda S_{\Lambda \perp} \rangle \langle X
P_\Lambda S_{\Lambda \perp}|J_\nu(0) |P\rangle  \ ,
\end{eqnarray}
where $l'$ is the momentum for the final state lepton,
$J^\mu$ is the quark electromagnetic current, and $X$
represents all other final-state hadrons other than the observed Hyperon $\Lambda$.

It is convenient to write the momentum of the virtual photon
in terms of the incoming and outgoing hadron momenta,
\begin{eqnarray}
q^\mu=q_t^\mu+\frac{q \cdot P_\Lambda}{P \cdot P_\Lambda}P^\mu +\frac{q \cdot P}{P \cdot P_\Lambda}P_\Lambda^\mu \ ,
\end{eqnarray}
where $q_t^\mu$ is transverse to the momentum of the initial and final hadrons:
$q_t^\mu P_\mu= q_t^\mu P_{\Lambda\mu}=0$.
Here $q_t^\mu$ is a space-like vector, and we
define $ \vec{q}^{\ 2}_\perp\equiv - q_t^2 $. In the hadron frame,
the final state hadron will have the momentum,
\begin{eqnarray}
P_\Lambda^\mu=\frac{x_B \vec{P}_{\Lambda\perp}^{\ 2}}{z_h Q^2} P^+ p^\mu
+ z_h \frac{Q^2}{2 x_B P^+} n^\mu + P^\mu_{\Lambda\perp}\ ,
\end{eqnarray}
where $P_{\Lambda\perp}$ is the Hyperon transverse momentum in the
hadron-frame, $P^+=1/\sqrt{2}(P^0+P^z)$, and we use the conventional
definition for light-cone vector $p^\mu,n^\mu$: $p=(1^+,0^-,0_\perp)$, $n=(0^+,1^-,0_\perp)$.
From the above definitions, we will find $\vec{q}^{\
2}_\perp=\vec{P}_{\Lambda\perp}^{\ 2}/{z_h^2}$.

We will calculate the hadronic tensor $W^{\mu\nu}$ at large transverse momentum
in perturbative QCD, by radiating a hard gluon in the final state. They are expressed in terms of
integrated parton distribution and fragmentation functions or the quark-gluon-antiquark correlations,
according to a collinear factorization~\cite{qiu-fac}. In the calculations, it is convenient to decompose
the hadronic tensor $W^{\mu \nu}$ in terms of individual tensors~\cite{Meng:1995yn,{Koike:2003zc}},
\begin{eqnarray}
W^{\mu \nu}=\sum^9_{i=1} V_i^{\mu \nu} W_i.
\end{eqnarray}
where the $W_i$ are structure functions, and can be projected out
from $W^{\mu \nu}$ by $W_i=W_{\alpha \beta} \tilde{V}_i^{\alpha
\beta}$ with the corresponding inverse tensors $\tilde{V}_i$. Both
$V_i$, $\tilde{V}_i$ can be constructed from four orthonormal basis
vectors~\cite{Meng:1995yn}: $T^\mu, X^\mu, Y^\mu ,Z^\mu $ with
normalization $T^\mu T_\mu=1,X^\mu X_\mu=Y^\mu Y_\mu=Z^\mu
Z_\mu=-1$. These four vectors can be further constructed by
$P^\mu$, $q^\mu$, $S_{\Lambda\perp}^\mu$, $q_t^\mu$.
In this paper, we choose a
special frame, where the $q_t^\mu$ is parallel to $X^\mu$,  and the
target proton and final state $\Lambda$ have spatial components only
in the $Z$ direction. In the small $q_t$ ($P_{\Lambda\perp}$)
region, we have checked that this frame leads to the same result as
that in the normal hadron frame.

As mentioned in the Introduction, in this paper, we are interested in
calculating the differential cross section in the intermediate transverse
momentum region, $\Lambda_{QCD}\ll P_{\Lambda\perp}\ll Q$. In the calculations,
we will utilize the power counting method, and only keep the leading
power contributions and neglect all higher order corrections in terms
of $P_{\Lambda\perp}/Q$. For the spin-average $\Lambda$ production
in SIDIS, the differential cross section will be identical to any other hadron
production process except we have to change the associated fragmentation
function for the Hyperon. This cross section in the above limit will be consistent
to the TMD factorization approach in the intermediate transverse momentum
region as what has been shown before, for example, in the pion production
in SIDIS~\cite{Ji:2006ub}.

For the $\Lambda$ polarization dependent cross section, we have two
separate contributions from the twist-three quark-gluon-antiquark
correlations in the parton distribution or fragmentation. In this
paper, we will only focus on the parton distribution part, whereas
that from the fragmentation part can follow accordingly. We also
note that these two contributions will have different azimuthal
angular dependence in SIDIS in the small transverse momentum limit.
For the contribution from the parton distribution part, following the
Qiu-Sterman formalism, the corresponding spin-dependent
hadronic tensor can be written as~\cite{qiu},
\begin{eqnarray}
W^{\mu\nu}=\int \frac{d^4k_1}{(2\pi)^4} \frac{d^4k_2}{(2\pi)^4}
T_\rho(k_1,k_2) H^{\mu\nu,\rho}(k_1,k_2,P_{\Lambda \perp},S_{\Lambda
\perp}) H_{1T}(z) \ , \label{eq1}
\end{eqnarray}
where $T$ and $H$ represent the twist-three
function and the partonic hard-scattering amplitude, respectively. These
two parts are connected by the two independent integrals over the momentum $k_1$
and $k_2$ that they share. In the above expansion, spinor and color
indices connecting the hard part and long-distance parts have
already been separated, which leads to the hard part $H(k_1,k_2,P_{\Lambda \perp},S_{\Lambda \perp})$
being contracted with $(1/2) \not\!\!p \gamma^\rho_\perp/(2\pi) $. The
transversity fragmentation $H_{1T}$ for $\Lambda$ production is
defined as
\begin{eqnarray}
H_{1T}(z)= \frac{1}{2z}\sum_{X} && \int \frac{dy^+}{4 \pi} e^{-i
P_{\Lambda}^- y^+/z}
\nonumber \\
&& \langle 0 | \psi(0^+) | P_{\Lambda} S_{\Lambda \perp},X \rangle
\langle P_{\Lambda} S_{\Lambda \perp},X | \bar{\psi}(y^+) S_{\perp
\mu} i \sigma^{\mu-} \gamma^5 |0 \rangle ,
\end{eqnarray}
where $X$ represents all other particles in the final state
except for $\Lambda$, $S_\perp^\mu$ is the transverse polarization
vector of the final state hadron. The next step is to perform a
collinear expansion of the expression:
\begin{eqnarray}
k_i^\mu=x_iP^\mu+k^\mu_{i,\perp} ,
\end{eqnarray}
where minus component has been neglected since it is beyond the
order in $k_{i,\perp}$ that we consider. The collinear expansion
enables us to reduce the four-dimensional integral to a integral
convolution in the light-cone momentum fractions of the initial
partons. Expanding $H^{\mu\nu}$ in the partonic momentum at
$k_1=x_1P$ and $k_2=x_2P$, we have
\begin{eqnarray}
H^{\mu\nu,\rho}(k_1,k_2,P_{\Lambda
\perp},S_{\Lambda\perp})&=&H^{\mu\nu,\rho}(x_1,x_2,P_{\Lambda
\perp},S_{\Lambda\perp})
\nonumber \\
&+&\frac{\partial H^{\mu\nu,\rho}}{\partial
k_1^\alpha}(x_1,x_2)(k_1-x_1P)^\alpha +\frac{\partial
H^{\mu\nu,\rho}}{\partial k_2^\alpha}(x_1,x_2)(k_2-x_2P)^\alpha+\dots
\end{eqnarray}
The above expansion allows us to integrate over three of the four
components of each of the loop momenta $k_i$, and the hadronic
tensor $W^{\mu\nu}$ will depend on the chiral-odd
spin-independent twist-three quark-gluon-antiquark correlation
function~\cite{qiu,yuji-l},
\begin{eqnarray}
T_F^{(\sigma)}(x_1,x_2)=\int \frac{dy_1^- dy_2^-}{4 \pi}
e^{iy_2^-(x_2-x_1)P^+- iy_1^-x_1P^+} \langle P| \bar{\psi}(y_1^-)
\sigma ^{+\mu} g F_{+\mu}(y_2^-) \psi(0) |P \rangle .
\end{eqnarray}
where the sums over color and spin indices are implicit, $|P \rangle
$ denotes the unpolarized proton state,  $\psi $ is the quark field,
and $ F_{+\mu}$ the gluon field tensor, and the gauge link has been
suppressed. Due to parity and time-reversal invariance, we have the
relation $T_F^{(\sigma)}(x_1,x_2)=T_F^{(\sigma)}(x_2,x_1)$.

Similar to the SSA in $\pi$ production in SIDIS, the strong
interaction phase necessary for having a non-vanishing $\Lambda$
transverse polarization arises from the interference between an
imaginary part of the partonic scattering amplitude with the extra
gluon and the real scattering amplitude without a gluon. The
imaginary part is due to the pole of the parton propagator
associated with the integration over the gluon momentum fraction
$x_g $ . Depending on which propagator's pole contributes, the
amplitude may get contributions from $ x_g=0$ (``soft-pole") and
$x_g\neq 0$ (``hard-pole" or ``soft-fermion-pole")~\cite{Ji:2006ub}.
The diagrams contributing to the $\Lambda$ polarization in SIDIS
will be the same as those calculated for the SSA in $\pi $
production. The only difference is that we have to replace the
Qiu-Sterman matrix element $T_F$ with the above unpolarized
quark-gluon-antiquark correlation function $T_F^{(\sigma)}$, and the
unpolarized fragmentation function $D(z)$ for $\pi$ with the
transversity fragmentation function for $\Lambda$. We further notice
that the soft-fermion-pole contribution is power suppressed in the
limit of $P_{\Lambda\perp}\ll Q$, similar to the SSA in $\pi$
production. In Fig.~1, we show some examples of the
``soft-pole" and ``hard-pole" diagrams. The calculations will be
similar to those in~\cite{qiu,Ji:2006ub,{new}}. We perform the calculation in
the covariant gauge. There are a total of eight diagrams
contributing to the soft-pole and twelve diagrams for the hard-pole
contributions. Since the calculation formalism has been established
well, we only give the final result and refer the reader to the
references for details.

\begin{figure}[t]
\begin{center}
\includegraphics[width=10cm]{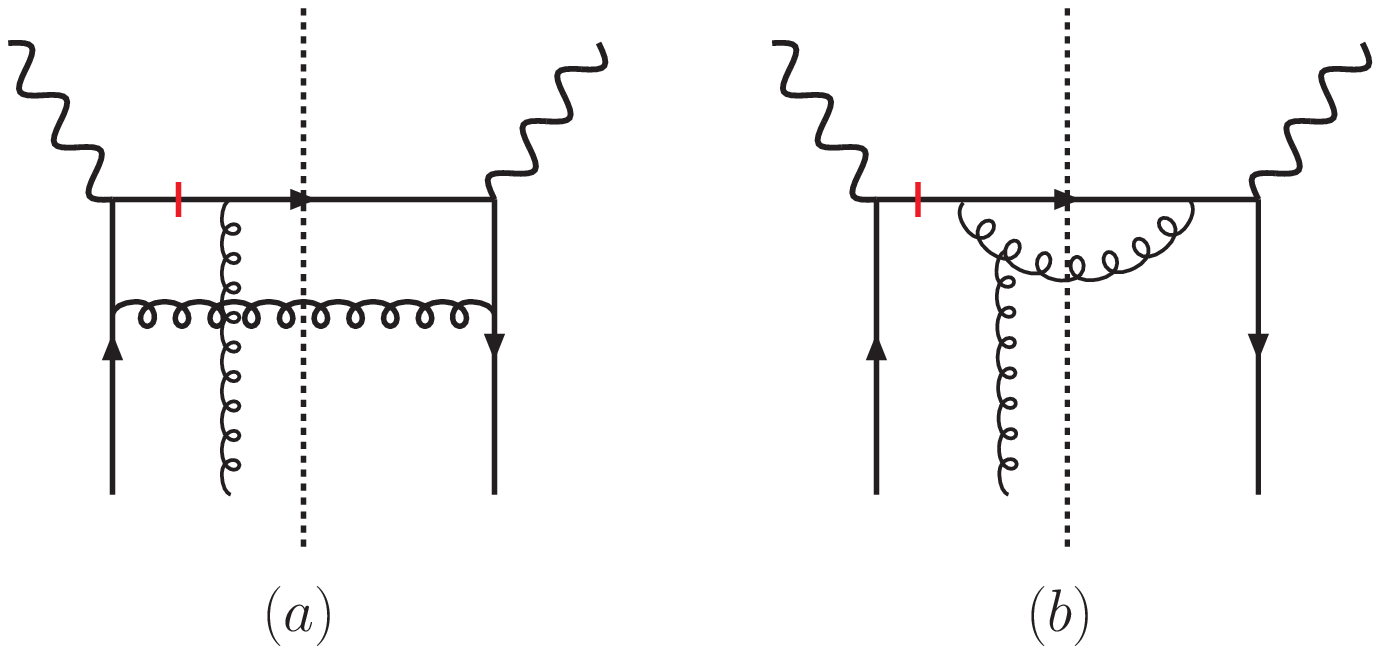}
\end{center}
\vskip -0.4cm \caption{The example diagrams for the soft pole (a)
and hard pole (b) contributions to $\Lambda$ transverse polarization
in semi-inclusive DIS process. The short bars indicate the pole
contribution from the propagators.} \label{fig1}
\end{figure}

We are interested to obtain the differential cross section in the limit of
$P_{\Lambda\perp}\ll Q$. In this limit, we further find that only $V_4$ and $V_9$ in Eq.~(6)
contribute in the leading power of $P_{\Lambda\perp}/Q$. $V_4$ and $V_9$ are defined as,
\begin{eqnarray}
V_4^{\mu\nu}&=&X^\mu X^\nu -Y^\mu Y^\nu\ ,\\
V_9^{\mu\nu}&=&X^\mu Y^\nu +X^\mu Y^\nu \ ,
\end{eqnarray}
and the associated $\tilde{V}_4$ and $\tilde{V}_9$ are given by,
\begin{eqnarray}
\tilde{V}_4^{\mu\nu}&=&\frac{1}{2} (X^\mu X^\nu- Y^\mu Y^\nu) \ , \\
\tilde{V}_9^{\mu\nu}&=&\frac{1}{2} (X^\mu Y^\nu + Y^\mu X^\nu)\  .
\end{eqnarray}
These two terms contribute the same to the differential cross sections
except the azimuthal angular dependence. The contribution from $V_4$
is proportional to $\cos(2\phi_\Lambda^l)\sin(\phi_s^l-\phi_\Lambda^l) $, where $\phi^l_\Lambda $
and $\phi_s^l$ are the azimuthal angles of the transverse momentum $P_{\Lambda\perp}$ and
the polarization vector $S_{\Lambda\perp}$ of $\Lambda$ relative
to the lepton scattering plane. On the other hand,
the $V_9$ contribution is proportional to $\sin(2\phi_\Lambda^l)\cos(\phi_s^l-\phi_\Lambda^l) $.
The total contributions from these two terms will be proportional to $\sin(\phi_\Lambda^l+\phi_s^l)$.

Summing up both soft-pole and hard-pole contributions from all the diagrams similar to
those in Fig.~1, we obtain the following $\Lambda$ polarization dependent differential
cross section from the contributions from $V_4$ and $V_9$,
\begin{eqnarray}
{\frac{d \sigma(S_{\Lambda\perp})}{dx_B dy dz_h d^2 \vec{P}_{\Lambda\perp}}}\Big|_{V_4+V_9}&=&
-\frac{4 \pi \alpha^2_{em} s}{Q^4} x_B(1-y) \sin(\phi^l_\Lambda+\phi^l_{s})
\frac{1}{z_h^2} \frac{\alpha_s}{2 \pi^2}
\nonumber \\
&&
\times\frac{2}{ |\vec{q}_\perp|} \int \frac{dxdz}{xz}
\delta \left ( \vec{q}_\perp^{\ 2}- \frac{Q^2(1-\xi)(1-\hat{\xi})}{\xi \hat{\xi}} \right )
\nonumber \\
&& \left \{ x \frac{\partial}{\partial x} T_F^{(\sigma)}(x,x)\frac{1}{2N_c}
\left[ \frac{(\xi+\hat{\xi}-1)}{\hat{\xi}(1-\hat{\xi})} \right] \right .\
\nonumber \\
&&\ - T_F^{(\sigma)}(x,x) \frac{1}{2N_c}
\left[ \frac{(\hat{\xi}-\xi^2+2\xi-1)}{(1-\xi)(1-\hat{\xi})\hat{\xi}}  \right]
\nonumber \\
&& \left .\ +T_F^{(\sigma)}(x,x_B)(\frac{1}{2N_c}+\hat{\xi} C_F)
\left[ \frac{1}{(1-\xi)(1-\hat{\xi})}    \right]  \right \}H_{1T}(z) \ ,
\end{eqnarray}
where $\xi=x_B/x$ and $\hat \xi=z_h/z$. In the above result,
the fist term in the bracket is the derivative term coming from the soft-gluon pole,
the second term is the non-derivation contribution from soft-gluon pole, and the third
is the hard-pole contribution whose derivative term vanishes.

In order to compare to the TMD factorization formalism,
we will extrapolate our results into the region of $ \Lambda_{QCD}\ll P_{h\perp}\ll Q $.
In doing the expansion, we only keep the terms leading in $P_{h\perp}/Q$,
and neglect all higher powers. For small $P_{h\perp}$, the delta function can be expanded as:
\begin{eqnarray}
\delta \left ( \vec{q}_\perp^{\ 2}- \frac{Q^2(1-\xi)(1-\hat{\xi})}{\xi \hat{\xi}} \right )
=\frac{\xi \hat{\xi}}{Q^2}
\left \{ \frac{\delta (\xi-1)}{(1-\hat{\xi})_+}
+\frac{\delta (\hat{\xi} -1)}{(1-\xi)_+}
+\delta (\xi-1) \delta(\hat{\xi}-1) \ln\frac{Q^2}{\vec{q}^{\ 2}_\perp} \right \}.
\end{eqnarray}
With this expansion,the spin-dependent cross section in the small $P_{h\perp}$ limit can be written as,
\begin{eqnarray}
\frac{d \sigma(S_{\Lambda \perp})}{dx_B dy dz_h d^2 \vec{P}_{\Lambda\perp}} \Big|_{P_{\Lambda\perp}\ll Q}&=&
-\frac{4 \pi \alpha^2_{em} s}{Q^4} x_B(1-y) \sin(\phi^l_h+\phi^l_{st})
\frac{1}{z_h^2} \frac{\alpha_s}{2 \pi^2}
\frac{1}{ |\vec{q}_\perp|^3}
\nonumber \\
&&
\int \frac{dxdz}{xz}H_{1T}(z)
\left \{ A \delta(\hat{\xi} -1)+ B \delta(\xi-1) \right \},
\end{eqnarray}
where
\begin{eqnarray}
&&A=\frac{1}{2 N_c} \left \{ \left [ x \frac{\partial}{\partial x} T_F^{(\sigma)}(x,x) \right ]
2 \xi+ T_F^{(\sigma)}(x,x)\frac{2 \xi(\xi-2)}{(1-\xi)_+} \right \}
+\frac{C_A}{2} T_F^{(\sigma)}(x,x_B)\frac{2}{(1-\xi)_+}
\nonumber \\
&&B=C_F T_F^{(\sigma)}(x,x) \left [ \frac{2 \hat{\xi}}{(1-\hat{\xi})_+}
+2 \delta (\hat{\xi} -1) \ln \frac{Q^2}{\vec{q}_\perp^{\ 2}} \right ] .
\end{eqnarray}
On the other side, the transverse-momentum-dependent factorization can be applied in the small
$P_{h\perp}\ll Q$. Therefore one expects the above result can be reproduced in this approach.
The $\Lambda$ polarization dependent cross section can be factorized as the following form~\cite{JiMaYu04,{ColMet04}}
\begin{eqnarray}
\frac{d \sigma(S_{\Lambda\perp})}{dx_B dy dz_h d^2 \vec{P}_{\Lambda\perp}}&=&
\frac{4 \pi \alpha^2_{em} s}{Q^4} x_B(1-y) \sin(\phi^l_\Lambda+\phi^l_{s})
\nonumber \\
&& \times \int \frac{k_\perp \cdot \hat{\vec{P}}_{\Lambda \perp}}{M} h_{1,DIS}^\perp(x,k_\perp)
H_{1T}(z,p_\perp) \left ( S(\lambda_\perp)\right )^{-1} H_{UUT}(Q^2) .
\end{eqnarray}
where $\hat{\vec{P}}_{\Lambda \perp}$ is the unit vector in direction of $\vec{P}_{\Lambda\perp}$,
$h_{1,DIS}^\perp$ is the TMD Boer-Mulders function for DIS process, and $H_{1T}$ the TMD transversity
fragmentation for $\Lambda$. $S(\lambda_\perp)$ and $ H_{UUT}(Q^2)$ are the soft factor
and hard factor, respectively.
The simple integral symbol represents a complicated integral:
$ \int=\int d^2 \vec{k}_\perp d^2 \vec{p}_\perp d^2 \vec{\lambda}_\perp
\delta^2(z \vec{k_\perp} +\vec{p}_\perp + \vec{\lambda}_\perp -\vec{P}_{\Lambda\perp} ).$
We have suppressed the sum over all flavors and factorization scale dependence in the
parton distribution function and fragmentation function.

When the $k_\perp$ is of the order of $\Lambda_{QCD}$ , the TMD dependent parton
distribution function are entirely non-perturbative objects. But in the region $\Lambda_{QCD} \ll k_\perp \ll Q$,
the TMD factorization still hold and at the same time $k_\perp$ dependent
parton distribution function $h_1^\perp$ can be calculated in terms of the twist-three
parton correlation function within the perturbative QCD.
This provides us a chance to make contact with the result from the collinear factorization formalism.
The perturbative calculation follow the similar procedure as that in~\cite{Ji:2006ub}. Finally, one obtains,
\begin{eqnarray}
h_{1,DIS}^\perp(x_B,k_\perp)=-\frac{\alpha_s}{2 \pi^2} \frac{M_p}{(\vec{k}_\perp^2)^2}
\int \frac{dx}{x} \left \{ A+ C_F T_F^{(\sigma)}(x,x) \delta(\xi-1)
\left ( \ln \frac{x_B^2 \zeta^2}{\vec{k}_\perp^2} -1 \right ) \right \} ,
\end{eqnarray}
where $A$ is given in Eq.(19), and $\xi=x_B/x$. We note that the native-time-reversal-odd
TMD Boer-Mulders function is process-dependent. The above result is for SIDIS. When
we apply the above to the Drell-Yan process, the Boer-Mulders function shall change the sign.
Similarly, for the TMD transversity fragmentation function, we have,
\begin{eqnarray}
H_{1T}(z_h,p_\perp)=\frac{\alpha_s}{2 \pi^2} \frac{1}{\vec{p}_\perp^2} C_F
\int \frac{dz}{z} H_{1T}(z) \left [ \frac{ 2 \hat{\xi}}{ (1-\hat{\xi})_+}
+ \delta(1- \hat{\xi}) \left ( \ln \frac{\hat{\zeta}^2}{\vec{p}_\perp^2} -1 \right ) \right ] ,
\end{eqnarray}
where $H_{1T}(z)$ is the integrated transversity quark fragmentation function defined in
Eq.~(8) and $\hat{\xi}=z_h/z$.

To obtain the final result, we let one of the transverse momentum
$\vec{k}_\perp,\vec{p}_\perp ,\vec{l}_\perp$ be of the order of $\vec{P}_\perp$
and the others are much smaller. After integrating the delta function, one has,
\begin{eqnarray}
\frac{d \sigma(S_{\Lambda\perp})}{dx_B dy dz_h d^2 \vec{P}_{\Lambda\perp}}&=&
-\frac{4 \pi \alpha^2_{em} s}{Q^4} x_B(1-y) \sin(\phi^l_\Lambda+\phi^l_{s})
\frac{ z_h}{|\vec{P}_{\Lambda\perp}|^3} \frac{\alpha_s}{2 \pi^2}
\nonumber \\
&& \times
\int \frac{dx dz}{x z}H_{1T}(z)
\left \{ A \delta(\hat{\xi} -1)+ B \delta(\xi-1) \right \},
\end{eqnarray}
where we have used the relation,
$T_F^{(\sigma)}(x,x)=-\int d^2 k_\perp \frac{|k_\perp|^2}{M_p}h^\perp_{1,DIS}(x,k_\perp^2)$ \cite{Boer:2003cm,Ma}.
Obviously, we reproduce the differential cross sections from the
collinear factorization calculation.

This clearly demonstrates that in the intermediate transverse momentum
region, the twist-three collinear factorization approach and the
TMD factorization approach provide a unique picture for the $\Lambda$ polarization
in the unpolarized semi-inclusive DIS process. This is because the observable
we calculated above is the leading contribution in the limit of $P_{\Lambda\perp}/Q$,
and the TMD factorization is valid~\cite{JiMaYu04,{ColMet04}}.

\section{$\Lambda$ polarization in unpolarized hadronic scattering}
The extension to the $\Lambda$ polarization in hadronic scattering is straightforward. The
diagrams will be similar to what have been calculated for the SSA in
inclusive hadron production in $pp_{\uparrow}\to \pi X$ collisions~\cite{new}. We need to
change the twist-three correlation function $T_F$ to our $T_F^{\sigma}$ and
couple to the transverse polarized fragmentation function for $\Lambda$. Again, we
will have contributions from quark-quark, and quark-gluon scattering channels.
Similar to what we have calculated in the last section, we will have both derivative and non-derivative
contributions. The derivative terms have been calculated in~\cite{yuji-l}. Using the
same method as that in~\cite{new}, we calculated the non-derivative terms.
To our surprise, from these calculations, we find that the non-derivative terms
vanish for $\Lambda$ polarization in hard partonic scattering processes. This
indicates that the compact formula~\cite{new} containing both derivative and non-derivative
contributions may not work in general, and shows a counter-example of the
derivation of the compact formula in~\cite{yuji-compact}. It will be interesting to
further investigate the reason for this observation.

\section{Conclusion }
In summary, in this paper, we studied the $\Lambda$ polarization in the unpolarized semi-inclusive DIS
and $pp$ collisions. In the SIDIS process, we compared the twist-three approach with the TMD
factorization approach and found that they are consistent with each other at the intermediate transverse
momentum region. The calculations in $pp$ collisions show an interesting pattern that there are
only derivative terms contributions and the non-derivative terms vanish.

We have also calculated the large transverse momentum behavior for the naive
time-reversal-odd Boer-Mulders quark distribution in the twist-three approach
from the quark-gluon-antiquark correlation function in unpolarized nucleon.
This result can be generalized to that in the Drell-Yan process. Because of T-odd
effects, we will have an opposite sign between these two processes,
\begin{equation}
h_{1,DY}^\perp(x,k_\perp)=-h_{1,DIS}^\perp(x,k_\perp) \ .
\end{equation}
This distribution has a number of important applications in the Drell-Yan lepton
azimuthal distribution in $pp$ scattering. For example, the
$\cos 2\phi$ angular distribution has contribution from
two Boer-Mulders functions from the incoming nucleons~\cite{boer-dy}. From our
calculations above, we shall be able to study the large transverse momentum
behavior for this $\cos 2\phi$ angular distribution. The extension to this will be carried out in
a separate publication.

This work was supported in part by the U.S. Department of Energy
under contract DE-AC02-05CH11231 and the National Natural Science
Foundation of China under the approval No. 10525523. We are grateful
to RIKEN, Brookhaven National Laboratory and the U.S. Department of
Energy (contract number DE-AC02-98CH10886) for providing the
facilities essential for the completion of this work. J.Z. is
partially supported by China Scholarship Council.


\begin{thebibliography}{99}

\bibitem{bunce}
%\cite{Bunce:1976yb}
%\bibitem{Bunce:1976yb}
  G.~Bunce {\it et al.},
  %``Lambda 0 Hyperon Polarization In Inclusive Production By 300-Gev Protons On
  %Beryllium,''
  Phys.\ Rev.\ Lett.\  {\bf 36}, 1113 (1976);
  %%CITATION = PRLTA,36,1113;%%
%\cite{Heller:1978ty}
%\bibitem{Heller:1978ty}
  K.~J.~Heller {\it et al.},
  %``Polarization Of Lambdas And Anti-Lambdas Produced By 400-Gev Protons,''
  Phys.\ Rev.\ Lett.\  {\bf 41}, 607 (1978)
  [Erratum-ibid.\  {\bf 45}, 1043 (1980)].
  %%CITATION = PRLTA,41,607;%%




%\cite{Anselmino:1994gn}
\bibitem{review}
  M.~Anselmino, A.~Efremov and E.~Leader,
  %``The theory and phenomenology of polarized deep inelastic scattering,''
  Phys.\ Rept.\  {\bf 261}, 1 (1995)
  [Erratum-ibid.\  {\bf 281}, 399 (1997)];
%  [arXiv:hep-ph/9501369].
  %%CITATION = PRPLC,261,1;%%
%\cite{Liang:2000gz}
%\bibitem{Liang:2000gz}
  Z.~T.~Liang and C.~Boros,
  %``Single spin asymmetries in inclusive high energy hadron hadron  collision
  %processes,''
  Int.\ J.\ Mod.\ Phys.\  A {\bf 15}, 927 (2000);
%  [arXiv:hep-ph/0001330].
  %%CITATION = IMPAE,A15,927;%%
%\cite{Barone:2001sp}
%\bibitem{Barone:2001sp}
  V.~Barone, A.~Drago and P.~G.~Ratcliffe,
  %``Transverse polarisation of quarks in hadrons,''
  Phys.\ Rept.\  {\bf 359}, 1 (2002).
%  [arXiv:hep-ph/0104283].
  %%CITATION = PRPLC,359,1;%%

%\cite{Qiu:1991pp}
\bibitem{qiu}
  J.~W.~Qiu and G.~Sterman,
  %``Single Transverse Spin Asymmetries,''
  Phys.\ Rev.\ Lett.\  {\bf 67}, 2264 (1991);
  %%CITATION = PRLTA,67,2264;%%
%\cite{Qiu:1991wg}
%\bibitem{Qiu:1991wg}
  %J.~w.~Qiu and G.~Sterman,
  %``Single transverse spin asymmetries in direct photon production,''
  Nucl.\ Phys.\  B {\bf 378}, 52 (1992).
  %%CITATION = NUPHA,B378,52;%%

%\bibitem{BroHwaSch02}
\bibitem{Brodsky:2002cx}
S.~J.~Brodsky, D.~S.~Hwang and I.~Schmidt,
%``Final-state interactions and single-spin asymmetries in
% semi-inclusive  deep inelastic scattering,''
Phys.\ Lett.\ B {\bf 530}, 99 (2002);
%%CITATION = HEP-PH 0201296;%%
%\cite{Brodsky:2002rv}
%\bibitem{Brodsky:2002rv}
%S.~J.~Brodsky, D.~S.~Hwang and I.~Schmidt,
%``Initial-state interactions and single-spin asymmetries in Drell-Yan
% processes,''
Nucl.\ Phys.\ B {\bf 642}, 344 (2002).
%%CITATION = HEP-PH 0206259;%%



%\bibitem{Col02}
%\cite{Collins:2002kn}
\bibitem{Collins:2002kn}
J.~C.~Collins,
%``Leading-twist single-transverse-spin asymmetries: Drell-Yan and
% deep-inelastic scattering,''
Phys.\ Lett.\ B {\bf 536}, 43 (2002).
%%CITATION = HEP-PH 0204004;%%

%\bibitem{BelJiYua02}
%\cite{Ji:2002aa}
\bibitem{Ji:2002aa}
X.~Ji and F.~Yuan,
%``Parton distributions in light-cone gauge: Where are the final-state
% interactions?,''
Phys.\ Lett.\ B {\bf 543}, 66 (2002);
%%CITATION = HEP-PH 0206057;%%
%\bibitem{ji2}
%\cite{Belitsky:2002sm}
%\bibitem{Belitsky:2002sm}
A.~V.~Belitsky, X.~Ji and F.~Yuan,
%``Final state interactions and gauge invariant parton
% distributions,''
Nucl.\ Phys.\ B {\bf 656}, 165 (2003).
%%CITATION = HEP-PH 0208038;%%


%\bibitem{BoeMulPij03}
%\cite{Boer:2003cm}
\bibitem{Boer:2003cm}
D.~Boer, P.~J.~Mulders and F.~Pijlman,
%``Universality of T-odd effects in single spin and azimuthal
% asymmetries,''
Nucl.\ Phys.\ B {\bf 667}, 201 (2003).
%%CITATION = HEP-PH 0303034;%%
%\cite{Anselmino:2004ky}

\bibitem{JiMaYu04}
%\cite{Ji:2004wu}
%\bibitem{Ji:2004wu}
  X.~Ji, J.~P.~Ma and F.~Yuan,
  %``QCD factorization for semi-inclusive deep-inelastic scattering at
% low
  %transverse momentum,''
  Phys.\ Rev.\ D {\bf 71}, 034005 (2005);
%  [arXiv:hep-ph/0404183];
  %%CITATION = HEP-PH 0404183;%%
%\bibitem{JiMaYu04p}
%\cite{Ji:2004xq}
%\bibitem{Ji:2004xq}
%X.~Ji, J.~P.~Ma and F.~Yuan,
%``QCD factorization for spin-dependent cross sections in DIS and
% Drell-Yan
%processes at low transverse momentum,''
Phys.\ Lett.\ B {\bf 597}, 299 (2004).
% [arXiv:hep-ph/0405085].
%%CITATION = HEP-PH 0405085;%%

%\cite{Collins:2004nx}
%\bibitem{Collins:2004nx}
\bibitem{ColMet04}
J.~C.~Collins and A.~Metz,
%``Universality of soft and collinear factors in hard-scattering
%factorization,''
Phys.\ Rev.\ Lett.\  {\bf 93}, 252001 (2004).
%[arXiv:hep-ph/0408249].
%%CITATION = HEP-PH 0408249;%%


%\cite{Efremov:1981sh}
\bibitem{Efremov}
  A.~V.~Efremov and O.~V.~Teryaev,
  %``On Spin Effects In Quantum Chromodynamics,''
  Sov.\ J.\ Nucl.\ Phys.\  {\bf 36}, 140 (1982)
  [Yad.\ Fiz.\  {\bf 36}, 242 (1982)];
  %%CITATION = SJNCA,36,140;%%
%\cite{Efremov:1984ip}
%\bibitem{Efremov:1984ip}
  A.~V.~Efremov and O.~V.~Teryaev,
  %``QCD Asymmetry And Polarized Hadron Structure Functions,''
  Phys.\ Lett.\ B {\bf 150}, 383 (1985).
  %%CITATION = PHLTA,B150,383;%%

%\cite{Kanazawa:2000hz}
\bibitem{Kanazawa:2000hz}
  Y.~Kanazawa and Y.~Koike,
  %``Chiral-odd contribution to single-transverse spin asymmetry in
% hadronic
  %pion production,''
  Phys.\ Lett.\ B {\bf 478}, 121 (2000);
  %[arXiv:hep-ph/0001021].
  %%CITATION = HEP-PH 0001021;%%
%\cite{Kanazawa:2000cx}
%\bibitem{Kanazawa:2000cx}
  %Y.~Kanazawa and Y.~Koike,
  %``Polarization in hadronic Lambda hyperon production and twist-3
% quark
  %distribution,''
  Phys.\ Rev.\ D {\bf 64}, 034019 (2001).
  %[arXiv:hep-ph/0012225].
  %%CITATION = HEP-PH 0012225;%%

%\cite{Kouvaris:2006zy}
\bibitem{new}
  C.~Kouvaris, J.~W.~Qiu, W.~Vogelsang and F.~Yuan,
  %``Single transverse-spin asymmetry in high transverse momentum pion
  %production in p p collisions,''
  Phys.\ Rev.\  D {\bf 74}, 114013 (2006).
%  [arXiv:hep-ph/0609238].
  %%CITATION = PHRVA,D74,114013;%%

%\cite{Eguchi:2006mc}
\bibitem{Eguchi:2006mc}
  H.~Eguchi, Y.~Koike and K.~Tanaka,
  %``Twist-3 formalism for single transverse spin asymmetry reexamined:
  %Semi-inclusive deep inelastic scattering,''
  Nucl.\ Phys.\  B {\bf 763}, 198 (2007).
%  [arXiv:hep-ph/0610314].
  %%CITATION = NUPHA,B763,198;%%

%\cite{Eguchi:2006qz}
\bibitem{Eguchi:2006qz}
  H.~Eguchi, Y.~Koike and K.~Tanaka,
  %``Single transverse spin asymmetry for large-p(T) pion production in
  %semi-inclusive deep inelastic scattering,''
  Nucl.\ Phys.\  B {\bf 752}, 1 (2006).
%  [arXiv:hep-ph/0604003].
  %%CITATION = NUPHA,B752,1;%%

%\cite{Koike:2007rq}
\bibitem{yuji-compact}
  Y.~Koike and K.~Tanaka,
  %``Universal Structure of Twist-3 Soft-Gluon-Pole Cross Sections for Single
  %Transverse-Spin Asymmetry,''
  Phys.\ Rev.\  D {\bf 76}, 011502 (2007).
%  [arXiv:hep-ph/0703169].
  %%CITATION = PHRVA,D76,011502;%%


%\cite{Sivers:1990fh}
\bibitem{Sivers:1990fh}
  D.~W.~Sivers,
  %``Hard scattering scaling laws for single spin production asymmetries,''
  Phys.\ Rev.\  D {\bf 43}, 261 (1991).
  %%CITATION = PHRVA,D43,261;%%

%\cite{Anselmino:1994tv}
\bibitem{Anselmino:1994tv}
  M.~Anselmino, M.~Boglione and F.~Murgia,
  %``Single spin asymmetry for p (polarized) p $\to$ pi X in perturbative QCD,''
  Phys.\ Lett.\  B {\bf 362}, 164 (1995);
%  [arXiv:hep-ph/9503290].
  %%CITATION = PHLTA,B362,164;%%
%\cite{Anselmino:1998yz}
%\bibitem{Anselmino:1998yz}
  M.~Anselmino and F.~Murgia,
  %``Single spin asymmetries in p(pol.) p and anti-p(pol.) p inclusive
  %processes,''
  Phys.\ Lett.\  B {\bf 442}, 470 (1998);
%  [arXiv:hep-ph/9808426].
  %%CITATION = PHLTA,B442,470;%%
%\cite{Anselmino:2004ky}
%\bibitem{Anselmino:2004ky}
  M.~Anselmino, M.~Boglione, U.~D'Alesio, E.~Leader and F.~Murgia,
  %``Parton intrinsic motion: Suppression of the Collins mechanism for
  %transverse single spin asymmetries in p(pol.) p --> pi X,''
  Phys.\ Rev.\  D {\bf 71}, 014002 (2005).
%  [arXiv:hep-ph/0408356].
  %%CITATION = PHRVA,D71,014002;%%


%\cite{Ji:2006ub}
\bibitem{Ji:2006ub}
  X.~Ji, J.~W.~Qiu, W.~Vogelsang and F.~Yuan,
  %``A unified picture for single transverse-spin asymmetries in hard
  %processes,''
  Phys.\ Rev.\ Lett.\  {\bf 97}, 082002 (2006);
%  [arXiv:hep-ph/0602239].
  %%CITATION = PRLTA,97,082002;%%
%\cite{Ji:2006vf}
%\bibitem{Ji:2006vf}
  %X.~Ji, J.~W.~Qiu, W.~Vogelsang and F.~Yuan,
  %``Single transverse-spin asymmetry in Drell-Yan production at large and
  %moderate transverse momentum,''
  Phys.\ Rev.\  D {\bf 73}, 094017 (2006);
%  [arXiv:hep-ph/0604023].
  %%CITATION = PHRVA,D73,094017;%%
%\cite{Ji:2006br}
%\bibitem{Ji:2006br}
 % X.~Ji, J.~W.~Qiu, W.~Vogelsang and F.~Yuan,
  %``Single-transverse spin asymmetry in semi-inclusive deep inelastic
  %scattering,''
  Phys.\ Lett.\  B {\bf 638}, 178 (2006);
%  [arXiv:hep-ph/0604128].
  %%CITATION = PHLTA,B638,178;%%
%\cite{Koike:2007dg}
%\bibitem{Koike:2007dg}
  Y.~Koike, W.~Vogelsang and F.~Yuan,
  %``On the Relation Between Mechanisms for Single-Transverse-Spin
  %Asymmetries,''
  arXiv:0711.0636 [hep-ph].
  %%CITATION = ARXIV:0711.0636;%%


%\cite{Dharmaratna:1996xd}
\bibitem{goldstein}
  W.~G.~D.~Dharmaratna and G.~R.~Goldstein,
  %``Single quark polarization in quantum chromodynamics subprocesses,''
  Phys.\ Rev.\  D {\bf 53}, 1073 (1996).
  %%CITATION = PHRVA,D53,1073;%%

%\cite{Liang:1997rt}
\bibitem{Liang:1997rt}
  Z.~T.~Liang and C.~Boros,
  %``Hyperon polarization and single spin left-right asymmetry in inclusive
  %production processes at high energies,''
  Phys.\ Rev.\ Lett.\  {\bf 79}, 3608 (1997);
%  [arXiv:hep-ph/9708488].
  %%CITATION = PRLTA,79,3608;%%
 H.~Dong and Z.~T.~Liang,
  %``Hyperon polarization in different inclusive production processes in
  %unpolarized high energy hadron hadron collisions,''
  Phys.\ Rev.\  D {\bf 70}, 014019 (2004)
  [arXiv:hep-ph/0403041].
  %%CITATION = PHRVA,D70,014019;%%

%\cite{Anselmino:2001js}
\bibitem{anselmino-l}
  M.~Anselmino, D.~Boer, U.~D'Alesio and F.~Murgia,
  %``Transverse Lambda polarization in semi-inclusive DIS,''
  Phys.\ Rev.\  D {\bf 65}, 114014 (2002).
%  [arXiv:hep-ph/0109186].
  %%CITATION = PHRVA,D65,114014;%%

%\cite{Kanazawa:2000cx}
\bibitem{yuji-l}
  Y.~Kanazawa and Y.~Koike,
  %``Polarization in hadronic Lambda hyperon production and twist-3 quark
  %distribution,''
  Phys.\ Rev.\  D {\bf 64}, 034019 (2001);
%  [arXiv:hep-ph/0012225];
  %%CITATION = PHRVA,D64,034019;%%
  %\cite{Koike:2002ar}
%\bibitem{Koike:2002ar}
  Y.~Koike,
  %``Hyperon polarization from unpolarized p p and e p collisions,''
  AIP Conf.\ Proc.\  {\bf 675}, 574 (2003).
%  [arXiv:hep-ph/0210434].
  %%CITATION = APCPC,675,574;%%


  %\cite{Boer:1997nt}
\bibitem{boer98}
  D.~Boer and P.~J.~Mulders,
  %``Time-reversal odd distribution functions in leptoproduction,''
  Phys.\ Rev.\  D {\bf 57}, 5780 (1998).
%  [arXiv:hep-ph/9711485].
  %%CITATION = PHRVA,D57,5780;%%

  %\cite{Boer:1999mm}
\bibitem{boer-dy}
  D.~Boer,
  %``Investigating the origins of transverse spin asymmetries at RHIC,''
  Phys.\ Rev.\  D {\bf 60}, 014012 (1999).
%  [arXiv:hep-ph/9902255].
  %%CITATION = PHRVA,D60,014012;%%

\bibitem{qiu-fac}
%\cite{Qiu:1990xxa}
%\bibitem{Qiu:1990xxa}
  J.~w.~Qiu and G.~Sterman,
  %``Power corrections in hadronic scattering. 1. Leading 1/Q**2 corrections to
  %the Drell-Yan cross-section,''
  Nucl.\ Phys.\  B {\bf 353}, 105 (1991);
  %%CITATION = NUPHA,B353,105;%%
%\cite{Qiu:1990xy}
%\bibitem{Qiu:1990xy}
 % J.~w.~Qiu and G.~Sterman,
  %``Power corrections to hadronic scattering. 2. Factorization,''
  Nucl.\ Phys.\  B {\bf 353}, 137 (1991).
  %%CITATION = NUPHA,B353,137;%%



%\cite{Meng:1995yn}
\bibitem{Meng:1995yn}
  R.~Meng, F.~I.~Olness and D.~E.~Soper,
  %``Semi-Inclusive Deeply Inelastic Scattering at Small q_T,''
  Phys.\ Rev.\  D {\bf 54}, 1919 (1996).
%  [arXiv:hep-ph/9511311].
  %%CITATION = PHRVA,D54,1919;%%

%\cite{Koike:2003zc}
\bibitem{Koike:2003zc}
  Y.~Koike and J.~Nagashima,
  %``Double spin asymmetries for large-p(T) hadron production in  semi-inclusive
  %DIS,''
  Nucl.\ Phys.\  B {\bf 660}, 269 (2003)
  [Erratum-ibid.\  B {\bf 742}, 312 (2006)].
%  [arXiv:hep-ph/0302061].
  %%CITATION = NUPHA,B660,269;%%




%\cite{Ma}
\bibitem{Ma}
  J.~P.~Ma and Q.~Wang,
  %``On unique predictions for single spin azimuthal asymmetry,''
  Eur.\ Phys.\ J.\  C {\bf 37}, 293 (2004).
%  [arXiv:hep-ph/0310245].
  %%CITATION = EPHJA,C37,293;%%


\end{thebibliography}
\end{document}